\documentclass[submission,copyright,creativecommons]{eptcs}

\usepackage{graphics}
\usepackage{graphicx}
\usepackage{listings}
\usepackage{todonotes}
\usepackage{amsmath}
\usepackage{amsthm}
\usepackage{amssymb,booktabs}
\usepackage{url}

\usepackage{pgf, tikz}
\usepackage{jiresch_in_simple}

\title{Towards a GPU-based implementation of interaction nets}
\author{Eugen Jiresch \thanks{The author was supported by the
Austrian Academy of Sciences (\"OAW) under grant no.\ 22932 and the Vienna PhD School of
Informatics.}
\institute{Institut f\"ur Computersprachen, Technische Universit\"at Wien, Favoritenstra{\ss}e 9-11, 1040 Wien, Austria}
\email{jiresch@logic.at}}
\def\authorrunning{E.\ Jiresch}
\def\titlerunning{Towards a GPU-based implementation of interaction nets}

\newcommand{\lwint}{$\;\stackrel{int}{\longrightarrow}\;$}
\newcommand{\lwcom}{$\;\stackrel{com}{\longrightarrow}\;$}
\newcommand{\lwcol}{$\;\stackrel{col}{\longrightarrow}\;$}
\newcommand{\lwsub}{$\;\stackrel{sub}{\longrightarrow}\;$}
\newcommand{\mlwint}{\stackrel{int}{\longrightarrow}}
\newcommand{\mlwcom}{\stackrel{com}{\longrightarrow}}
\newcommand{\mlwcol}{\stackrel{col}{\longrightarrow}}
\newcommand{\mlwsub}{\stackrel{sub}{\longrightarrow}}

\begin{document}

\lstset{language=C++,
		basicstyle=\ttfamily,
		mathescape=true
		}

\maketitle

\begin{abstract}
We present \emph{ingpu}, a GPU-based evaluator for interaction nets that
heavily utilizes their potential for parallel evaluation.
We discuss advantages and challenges of the ongoing implementation of
\emph{ingpu} and compare its performance to existing interaction nets
evaluators.
\end{abstract}


\section{Introduction}
\label{sec:introduction}

\emph{Interaction nets} are a model of computation based on graph
rewriting.
They enjoy several useful properties
which makes them a promising candidate for a future functional programming
language. In particular, reducible expressions in a net can be evaluated in any
order, even in parallel. This makes an implementation of interaction nets on a
multicore architecture attractive.

However, the amount of parallelism in an interaction net is highly dynamic,
and depends on the particular program and even runtime values. At any point
during a computation, the number of expressions that can be evaluated in
parallel can vary between dozens and hundreds of thousands. There is currently
no implementation of interaction nets that leverages their full parallelism
potential.

In recent years, a trend towards using graphics processing units (GPUs) for
general purpose computations has emerged. Due to the increasing programability
of GPUs and general purpose APIs (CUDA, OpenCL), the parallel processing
power of graphics cards may be used for many kinds of problems, from simulation
of physical phenomena to cracking passwords. In this paper, we investigate the
parallel evaluation of interaction nets using GPUs. While the GPU model of parallelism seems to fit interaction nets well, several factors make an implementation a
non-trivial task. We argue that these factors can be overcome, thus enabling an
efficient evaluation of interaction nets. Furthermore, we describe our ongoing
prototype implementation.
Our contributions can be summarized as follows:
\begin{itemize}
  \item We describe \emph{ingpu}, the first interaction nets evaluator that
  runs almost entirely on the GPU, and heavily utilizes their potential for
  parallel evaluation.%
  \footnote{The source code is available at
  \url{https://github.com/euschn/ingpu} .}
  \item We describe the underlying algorithm and show its correctness.
  \item We present benchmark results of the ongoing implementation, and discuss
  possible solutions to overcome current performance drawbacks.
\end{itemize}

In the
next section, we recall the main notions of interaction nets and the lightweight calculus, which is the basis for our implementation. Section
\ref{sec:parallel-evaluation-with-cuda} describes the current status of our
prototype implementation using the CUDA \emph{Thrust} library. We follow this
description with benchmark results and discuss possible performance
improvements in Section \ref{sec:benchmarks}.
Finally, we discuss related work and conclude in Section \ref{sec:discussion}.


\section{Preliminaries}
\label{sec:prelims}

\subsection{Interaction nets}
\label{sec:interaction_nets}
We now recap the main notions of interaction nets and the
lightweight interaction calculus. \emph{Interaction nets} were first introduced in
\cite{lafont_interaction_1990}.
A \emph{net} is a graph consisting of \emph{agents} (labeled nodes) and
\emph{ports} (edges). Every agent has exactly one \emph{principal} port
(denoted by an arrow), all other ports are called \emph{auxiliary} ports. The
number of auxiliary ports is the \emph{arity} of the agent. Agent labels
denote data or function symbols, and are assigned a fixed arity. Computation is
modeled by rewriting the graph, which is based on \emph{interaction rules}.
These rules apply to two
nodes which are connected by
their \emph{principal ports}
(indicated by the arrows), forming an \emph{active pair} (or \emph{redex}).
For example, the following
rules model the addition of natural numbers (encoded by 0 and a successor function $S$):

\begin{tikzpicture}

\matrix[row sep=4mm, column sep=2mm] {
	\node			[]	{(1)};
	&
	& \node (res) {}; \\
	&
	& \node (min) [agent_small] {$+$};	& \\
	&
	\node (x) [agent_small] {$0$};	& & 	
	\node (y)  {\small{$y$}}; \\
	&
	\node (xp) {}; & &
	\node (yp) {}; \\	
};
	\link{min}{res}
	\activepair{min}{x}
	\link{y}{min}

\node (is) [right of= y, above=10pt] {\large{$\Rightarrow$}};

\matrix[right of=is, row sep=4mm, column sep=2mm, yshift=8pt] {
	& \node (res) {}; \\
	& \node (s) {};	& \\
	\\
	\node (x) {}; & &
	\node (yp) {\small{$y$}}; \\	
};

	\link[out=90, in=270]{yp}{res}

\matrix[right of= is,, xshift =80pt, yshift=8pt, row sep=3mm, column
sep=1.5mm] {
	\node			[]	{(2)};
	&
	& \node (res) {}; \\
	&
	& \node (min) [agent_small] {$+$};	& \\
	&
	\node (x) [agent_small] {S};	& & 	
	\node (y) 		  {\small{$y$}}; \\
	&
	\node (xp) {\small{$x$}};\\	
};
	\draw (min) to (res);
	\activepair{x}{min}
	\link{y}{min}
	\draw (x) to (xp);

\node (is) [right of= y, above=10pt] {\large{$\Rightarrow$}};

\matrix[right of=is, row sep=3mm, column sep=1.5mm] {
	& \node (res) {}; \\
	& \node (plus) [agent_small] {S}; \\
	& \node (s) [agent_small] {$+$};	& \\
	\node (x) {\small{$x$}}; & &
	\node (y) {\small{$y$}}; \\	
};
	\pp{plus}{res}
	\link{s}{plus}
	\pp{s}{x}
	\link{s}{y}
	
\end{tikzpicture}

This simple system allows for parallel evaluation of programs because by
definition active pairs are completely independent. Reducing a
pair cannot change, destroy or duplicate another one. Furthermore, any order of evaluation yields the same
result. Active pairs may even be reduced in parallel. This is due to the
following property.

\newtheorem{def:ucr}{Proposition}[subsection]

\begin{def:ucr} [\textbf{Uniform Confluence} \cite{lafont_interaction_1990}]
Let $P$ be an interaction net, and
$\Rightarrow$ the reduction relation induced by a set of rules $R$. Then the
following holds: if $N \Rightarrow P$ and $N \Rightarrow Q$ where $P \neq Q$,
then there exists some $R$ such that $P \Rightarrow R \Leftarrow Q$.%
\end{def:ucr}

\newtheorem{exp:addition_reduction}[def:ucr]{Example}

\begin{exp:addition_reduction}
Consider the interaction rules for addition of natural numbers. We can model the
evaluation of the term $1+(0+1)$ with the following reduction in interaction
nets (the active pair evaluated in each step is bold):

\tikzset{ every node/.style={scale=0.7} }

\begin{tikzpicture}

\matrix[row sep=3mm, column sep=3mm]{
	&
	\iface{r}
	\\
	&
	\agent{p1}{$+$} & &
	\agent{p2}{$+$}
	\\
	\agent{s1}{S} & &
	\agent{z2}{$0$} & &
	\agent{s2}{S} &
	\is
	\\
	\agent{z1}{$0$} & & & &
	\agent{z3}{$0$}
	\\
};
\link{p1}{r}
\link[out=300, in=90]{p1}{p2}
\ap[ultra thick]{p1}{s1}
\ppr{z1}{s1}
\ap{z2}{p2}
\ppl{s2}{p2}
\ppr{z3}{s2}

\end{tikzpicture}
\begin{tikzpicture}

\matrix[row sep=3mm, column sep=3mm]{
	&
	\iface{r}
	\\
	&
	\agent{s1}{S}
	\\
	&
	\agent{p1}{$+$} & &
	\agent{p2}{$+$}
	\\
	\agent{z1}{$0$} & &
	\agent{z2}{$0$} & &
	\agent{s2}{S} &
	\is
	\\
	& & & &
	\agent{z3}{$0$}
	\\
};
\ppr{s1}{r}
\link[out=300, in=90]{p1}{p2}
\link{p1}{s1}
\ap[ultra thick]{z1}{p1}
\ap{z2}{p2}
\ppl{s2}{p2}
\ppr{z3}{s2}

\end{tikzpicture}
\begin{tikzpicture}

\matrix[row sep=3mm, column sep=3mm]{
	&
	\iface{r}
	\\
	&
	\agent{s1}{S}
	\\
	& &
	\agent{p2}{$+$}
	\\
	&
	\agent{z2}{$0$} & &
	\agent{s2}{S} &
	\is
	\\
	& & &
	\agent{z3}{$0$}
	\\
};
\ppr{s1}{r}
\link[out=270, in=90]{s1}{p2}
\ap[ultra thick]{z2}{p2}
\ppl{s2}{p2}
\ppr{z3}{s2}
\end{tikzpicture}
\begin{tikzpicture}
\matrix[row sep=3mm, column sep=4mm] {
	\iface{r}
	\\
	\agent{s1}{S}
	\\
	\agent{s2}{S}
	\\
	\agent{z}{0}
	\\
};
\ppr{s1}{r}
\ppr{s2}{s1}
\ppr{z}{s2}
\end{tikzpicture}

\tikzset{ every node/.style={scale=1} }

After three steps, we reach the expected result $S(S(0))$, corresponding to $2$.
Note that each step only reduces one active pair. However, the third step
(reduction of $+$ and $0$) could be applied in parallel with either step one or
two, yielding the same result.
\end{exp:addition_reduction}

\paragraph{\textbf{Remark}}
This example illustrates the dynamics of the parallelism of interaction nets:
depending on the state of the computation, one or two active pairs
exist at the same time, and can hence be evaluated in parallel. For particular
sets of rules and input nets, the number of concurrent active pairs can get
as high as hundreds of thousands. On the other hand, some interaction nets may be inherently sequential.

\subsection{The lightweight calculus}
\label{sec:lightweight_calculus}

The \emph{lightweight calculus} \cite{DBLP:journals/eceasst/HassanMS10} is a textual
representation of interaction nets, providing the basis for our
implementation. It handles application of rules as well as rewiring and connecting of ports and
agents. It uses the following ingredients:
\begin{description}
  \item[Symbols $\Sigma$] representing agents, denoted by $\alpha,\beta,\gamma$.
  \item[Names $\mathrm{N}$] representing ports, denoted by
  $x,y,z,x_1,y_1,z_1,\ldots$ . We denote sequences of names by
  $\overline{x},\overline{y},\overline{z}$.
  \item[Terms $\mathrm{T}$]	being either names or symbols with a number of
  subterms, corresponding to the agent's arity: $t = x \; | \; \alpha(t_1,\ldots,t_n)$ .
  Terms are denoted by $s,t,u$, while  $\overline{s},\overline{t},\overline{u}$
  denote sequences of terms.
  \item[Equations $E$] denoted by $t=s$ where $t,s$ are terms, representing
  connections in a net. Note that $t = s$ is equivalent to $s = t$.
  We use $\Delta,\Theta$ to denote multisets of equations.
  \item[Configurations $C$] representing a net by $\langle \overline{t}\;|\; \Delta
  \rangle$. $\overline{t}$ is the interface of the net, i.e., its ports that are not
  connected to an agent. All names in a configuration occur at most twice. Names
  that occur twice are called \emph{bound}.
  \item[Interaction Rules $\mathrm{R}$] denoted by $\alpha(\overline{x}) =
  \beta(\overline{y}) \longrightarrow \Theta$. $\alpha,\beta$ is the active pair of the
  left-hand side (LHS) of the rule and the set of equations $\Theta$ represents
  the right-hand side (RHS).
\end{description}
Rewriting a net is modeled by applying four \emph{reduction rules} to a
configuration:
\begin{description}
  \item[Communication:] $\langle\; \bar{t} \;|\; x=t,x=u,\Delta \rangle
  \mlwcom \ \langle\; \bar{t} \;|\; t=u,\Delta \rangle $
  \item[Substitution:] $\langle\; \bar{t} \;|\; x=t,u=s,\Delta \rangle
  \mlwsub \ \langle\; \bar{t} \;|\; u[t/x]=s,\Delta \rangle $, where
  $u$ is not a variable.
  \item[Collect] $\langle\; \bar{t} \;|\; x=t,\Delta \rangle
  \mlwcol \ \langle\; \bar{t}[t/x] \;|\; \Delta \rangle $, where $x$
  occurs in $\bar{t}$.
  \item[Interaction] $\langle\; \bar{t}  \;|\;  \alpha(\bar{t_1}) =
  \beta(\bar{t_2}),\Delta\rangle \mlwint \langle\; \bar{t} \;|\;
  \Theta',\Delta\rangle$, where $\alpha(\bar{s_1}) =
  \beta(\bar{u_2}) \rightarrow \Theta \; \in \mathrm{R}$. $\Theta'$ denotes
  $\Theta$ where all bound variables in $\Theta$ are replaced by fresh variables
  and $\bar{s},\bar{u}$ are replaced by $\bar{t_1}, \bar{t_2}$.
\end{description}
Intuitively, \lwint replaces an active pair by the RHS of the corresponding
rule. The other three reduction rules substitute variables, which corresponds to
resolving a connection between two agents.

\newtheorem{exp:addition_calculus}{Example}

\begin{exp:addition_calculus}
The rules for addition presented in Section \ref{sec:interaction_nets}
are expressed in the lightweight calculus as follows:
\begin{align}
+(y,r) = S(x) & \;\;\longrightarrow\;\;  +(y, w) = x, \; r = S(w) \\
+(y,r) = Z  &  \;\;\longrightarrow\;\;  r = y
\end{align}
The following reduction calculates $1 + 0$:

\begin{align*}
 	\langle\; r  \;|\;  +(r,0) = S(0)\;\rangle & \mlwint &
 	\langle\; r  \;|\; r=S(x), +(x,0) = 0 \;\rangle
 	&\mlwcol& \langle\; S(x)  \;|\;  +(x,0) = 0\;\rangle \\
 	&\mlwint& \langle\; S(x)  \;|\;  x = 0 \;\rangle
 	&\mlwcol& \langle\; S(0)  \;|\;  \;\rangle
\end{align*}

\end{exp:addition_calculus}

It is important to note that
substitutions done by \lwsub and \lwcol never yield a new active pair/equation.
 This means that we can
reach a normal form of a net (i.e., free of active pairs) by using only \lwint
and \lwcom rules:

\newtheorem{thm_intcom}{Theorem}
\begin{thm_intcom} [\cite{DBLP:journals/eceasst/HassanMS10}]
\label{thm_intcom}
If $C_1 \longrightarrow^* C_2$ then there is a configuration $C$ such that
$C_1 \longrightarrow^* C \mlwsub^* . \mlwcol^* C_2$, and $C_1$ is reduced to
$C$ applying only communication and interaction rules.
\end{thm_intcom}

This theorem can be interpreted as follows. In the lightweight calculus, the
lion's share of the computation is done by \lwint and \lwcom. The former
reduces expressions (i.e., equations/active pairs) by replacing them with the
RHS of the corresponding interaction rule. The latter generates new active
equations that can be reduced by \lwint. The collect and substitution steps are in a sense
cosmetic: both resolve variables in order to provide a better readable form of
the result net. They perform no ``actual'' computation, i.e., rewriting of the
graph represented by the set of equations. Hence, all \lwsub and \lwcol steps
can be pushed to the end of the computation.


\section{Parallel evaluation of interaction nets in CUDA/Thrust}
\label{sec:parallel-evaluation-with-cuda}

In this section, we discuss the ongoing implementation of our GPU-based
interaction nets evaluator \emph{ingpu}. First, we give a quick introduction to
CUDA/\emph{Thrust} and motivate a GPU-based approach. We then describe the main
components of the evaluator, the interaction and the communication phase.

\subsection{CUDA and Thrust}
The tool \emph{ingpu} is written in C++ and CUDA. The latter is a language for
GPU-based, general-purpose
computation on NVIDIA graphics cards.
The general flow of a program using the
GPU for data-parallel computation is as follows: an array of input data sets is
copied from the main memory (known as \emph{host} memory) to the memory of the
GPU (also referred to as \emph{device}). A function (the \emph{kernel}) is
executed on the GPU in parallel on each individual data set. Finally, the array
of results is copied back to main memory.

In general, implementing an algorithm on a GPU efficiently requires
a considerable amount of low-level decisions: factors such as size of data
structures, number of threads and thread block size can greatly influence
performance. Fortunately, version 4.0 of CUDA introduced the \emph{Thrust}
library \cite{Thrust}, which features high-level constructs for efficiently
performing parallel computations. For example, \emph{Thrust} provides the
\lstinline;transform(); function, which is similar to \lstinline;map; in
functional programming:
%
%
%
\begin{lstlisting}
// allocate three device_vectors with 10 elements
thrust::device_vector<int> X(10), Y(10);
// compute Y = -X
thrust::transform(X.begin(), X.end(), Y.begin(),
   thrust::negate<int>());
\end{lstlisting}
\emph{Thrust} is obviously inspired by the C++ STL: \lstinline;device_vector; is a
generic, resizable container residing in GPU memory. The arguments of
\lstinline;transform; are an input vector \lstinline;X;, an output vector
\lstinline;Y; and a function object, here a built-in function that negates
integers.

Other functions supplied by \emph{Thrust} include parallel sorting, filtering
and reduction. Reductions compute a single value based on an input list, e.g.,
the sum of its elements:

\begin{lstlisting}
//sum the vector X, with initial fold value 0
sum = thrust::reduce(X.begin(), X.end(), 0, thrust::plus<int>());
\end{lstlisting}
%

These functions are a
convenient way to write parallel programs without the need for low-level
tweaking. Our interaction nets evaluator \emph{ingpu} is completely based on the
\emph{Thrust} library.

\subsection{Motivation and challenges}
\label{sec:motivation_and_challenges}
Why does a GPU-based implementation of interaction nets make sense in the first
place? Several reasons can be given: first, the SIMD (Single Instruction, Multiple
Data) model of GPUs is similar to the idea behind interaction nets. Several
independent data sets are processed in parallel using the same
instructions/program. This is analogous to reducing several active pairs with a
common set of interaction rules. Second, the reduction of a single active pair is a
fairly small computation, consisting only of a few lines of code. GPUs are
optimized for running thousands of threads executing such small programs.
Additionally, the number of active pairs existing at the same time may vary
greatly through the execution of a program. An interaction nets evaluator should
be able to dynamically and transparently scale this potential parallelism to the
many-core hardware. Again, GPUs are a promising platform to achieve this.

However, the implementation of interaction nets on a GPU is a non-trivial task.
In particular, it poses the following challenges:
\begin{description}
  \item[Maintaining the net structure] While active pairs can be reduced in
  parallel, they are not completely independent: they are connected in a net,
  and resolving these connections (via \lwcom) is needed to generate further
  active pairs. Unfortunately, the choice of data structures in GPU memory is very limited
  (essentially just arrays). Moreover, typical GPU programs are most efficient
  for algorithms with regular data access (for example, dense matrix
  multiplication). This means that it is difficult to efficiently
  represent the irregular graph structure of an interaction net.
  \item[Varying output size of a reduction]
  In general, the RHS of an interaction rule may be an arbitrarily large net.
  This implies that the result of a reduction of an active pair may vary in
  size, depending on the rule being used. Analogously, the number of new
  equations generated by one interaction rule in the lightweight calculus
  varies.
  Reducing one active pair may yield
  an arbitrarily large net, or any number of equations in the lightweight
  calculus.
  GPU-based algorithms usually have a fixed output size for every
  input.
\end{description}
Solving these challenges is by far not completed. In fact, we shall see that
dealing with these issues results in the performance deficits of the current
implementation, cf. Section \ref{sec:benchmarks}.

\subsection{Overview of the implementation}
\label{sec:overview_of_the_implementation}
In this subsection, we describe the basic concept behind \emph{ingpu}.
We represent agents and variables by a
\emph{unique id}, a \emph{name}, an \emph{arity} and a \emph{list of ids} of the agents
connected to an agent's auxiliary ports.
Currently, agents and variables are simply distinguished by upper and lower
case names. In fact, the name of a variable is not important: its identification
by the unique id is sufficient for all computations.
Naturally, we represent equations as pairs of agents. A configuration is simply
a vector of equations - we leave the interface of the net implicit.

The basic control flow of our evaluator is simple. Recall the essence of
Theorem \ref{thm_intcom}: to reach a result net that is free from active pairs,
it is sufficient to apply the interaction and communication rules to the set of
equations. Therefore, \emph{ingpu} performs parallel interaction and parallel
communication in a loop until no more active pairs exist.
For the remainder of this section, we shall describe the implementation of the
data-parallel versions of \lwint and \lwcom.

\subsection{Parallel interaction}

The \emph{interaction step} \lwint can be parallelized in a straightforward
way. Clearly, \lwint fits the SIMD model very well: the same program (i.e., the set
of interaction rules) is applied to each active pair, and replacing a pair is
completely independent of all others.

The problematic part about the \lwint step is the varying output size of an
active pair reduction, which we mentioned in Section
\ref{sec:motivation_and_challenges}.
In general, a rule RHS may contain an arbitrary number of equations.
Unfortunately, \emph{Thrust}'s algorithms can only return a fixed number of
results per individual input.
This means that it is not feasible to return a dynamically-sized list of
equations as the result of an interaction.
For the time being, we have solved this problem in a pragmatic way by
setting a maximum number \emph{n} of equations per RHS.
 Applying
any interaction rule must then yield exactly $n$ equations. Between zero and $m
\leq n$ equations represent the actual result. The remaining $n - m$ equations are
\emph{dummies}, resulting in a fixed result size for each application of the
kernel.

\begin{figure}[htpb]
\begin{center}
  \includegraphics[scale=0.55]{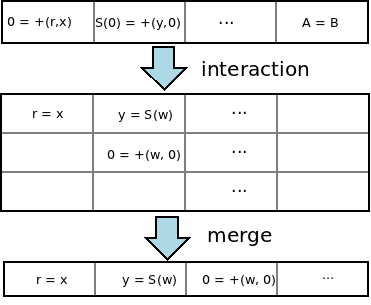}
  \caption{The interaction step}
  \label{fig:int_step}
\end{center}
\end{figure}

Obviously, the dummy equations must be filtered in a subsequent computation
step. Fortunately, \emph{Thrust} provides the \lstinline;remove(); function,
which is a parallel version of Haskell's \lstinline;filter;. Figure
\ref{fig:int_step} illustrates the full interaction step: every equation in the input array that
represents an active pair is reduced, i.e., the $n$ result slots ($n=3$ in the
figure) are filled with the equations of the RHS of the corresponding rule and
dummy equations (empty slot). Afterwards, the dummy equations are removed and
the resulting equations are merged into a single array.

This approach is straightforward, but has performance drawbacks. The filter and
merge operations on the result arrays are up to several hundred times slower
than the actual parallel interaction step. This contributes to the current
inefficiency of \emph{ingpu}, which we shall discuss in detail in Section
\ref{sec:benchmarks}.

\subsection{Parallel communication}
\label{sec:parallel-communication}
After the interaction phase above has completed, we apply an algorithm
corresponding to the \emph{communication} rule to generate new active pairs.
Recall the mechanics of \lwcom: communication needs to find two equations
$\{ x=t, x=s \}$ where $x$ is a variable and $s,t$ are terms and merge them to a
single equation $\{ s=t \}$.
Let us call two such equations sharing a variable \emph{communication-eligible}.
 This is harder to parallelize than \lwint: we need
to find eligible pairs of equations first before we reduce them. Unfortunately, this is where we
run into a problem mentioned in Section \ref{sec:motivation_and_challenges}: our
net is only represented as an array of independent equations. There is no
additional pointer structure between them to represent connections (cf. the
\texttt{Agent} structure in Section \ref{sec:overview_of_the_implementation}).
Therefore, we currently perform the following (inefficient) procedure to find
communication-eligible pairs of equations: we sort all equations of
the pattern $x=t$ by the id of the variable $x$.
This places communication-eligible pairs of equations next to each other in the
array (since the same variable occurs in them). We can now proceed to merge
these pairs into single equations. We do this by using \emph{Thrust}'s
\lstinline{reduce_by_key()}: this function essentially works like Haskell's
\texttt{fold} with an added predicate function. Only adjacent array elements
that satisfy the predicate are folded. For example, consider a list of numbers
$\{2,0,3,3,3,7,5,5\}$: calling \lstinline{reduce_by_key()} on this list with
addition as folding function and equality as predicate yields the result
$\{2,0,9,7,10\}$.

Hence, we use having a common variable as \lstinline;reduce_by_key();'s
predicate. This way, we merge communication-eligible pairs and leave all other
equations untouched. The communication step can be summed up by the following
pseudo-code:

\begin{lstlisting}
let E be the array of all equations of shape x=t
sort E by the id of left part of each equation
reduce_by_key(E) with predicate p and reduce functor r
  where
  p (x,t) (y,s) = x==y
  r (x,t) (y,s) = (s,t)
\end{lstlisting}

\emph{Thrust} provides efficient implementations of parallel sorting algorithms,
but sorting and reducing the equations still represents a major performance
bottleneck for large inputs. We shall discuss this issue in
detail in Section \ref{sec:benchmarks}. For the remainder of this section, we
shall show that our algorithm is correct.

\subsection{Correctness of the algorithm}
\label{sec:correctness}
The communication algorithm seems fairly straightforward, but we have not
discussed one important point: what if an equation consists of two variables,
e.g. $x=t$ where $t$ is a variable? Which one is used for comparison in the
sorting of the equations? Surprisingly, it does not matter for our algorithm.
We shall now show that it is correct in the sense that any possible active pair
that can be generated by \lwcom in the lightweight calculus will also be
generated by \emph{ingpu}.

First however, we highlight that some active pairs cannot be generated by
the communication algorithm described above. Consider a set of equations
$\{A=x,x=y,y=B\}$. Clearly, both $x$ and $y$ can be resolved, yielding the
active pair $A=B$. However, we need to perform two consecutive \lwcom steps,
first eliminating $x$ and then $y$ or the other way around.%
\footnote{The order of reduction steps in the lightweight calculus does not
influence the result. This is shown for the interaction calculus in
\cite{DBLP:conf/ppdp/FernandezM99}, which is the basis for the lightweight
calculus.}
The communication algorithm above would only perform one of these
communications, depending on whether $x$ or $y$ was used for sorting. The result
would be $\{A=y,y=B\}$ (if x was reduced), which cannot be used in the
interaction phase. A second pass of the communication algorithm would be
necessary to yield the active equation $A = B$. The following pseudo-code of the
evaluation function reflects this:
\begin{lstlisting}
equation_list evaluate(equation_list L) {
   transfer L to device (GPU) memory
   do {
      perform interaction as in Section 3.4
      perform communication as in Section 3.5
   }
   while (at least one interaction or one communication
      has occurred in the previous loop)
   transfer L back to host (CPU) memory
   return L
}
\end{lstlisting}

This way, if the communication algorithm does not yield any active pairs, it is
performed again until it either generates new active pairs or no longer performs
any \lwcom steps. In the latter case, the program terminates. We can show that
the algorithm is correct in the sense that it generates all possible active
pairs and evaluates them.

\newtheorem{prop:algorithm_correctness}[subsection]{Proposition}

\begin{prop:algorithm_correctness} [Correctness of \texttt{evaluate()}] Let $E$
be a set of equations such that there is an interaction rule for every active
pair. Then, $R=$ \lstinline;evaluate(;$E$\texttt{)} contains no active pairs,
and no further active pairs can be generated by applying \lwcom in $R$.
\end{prop:algorithm_correctness}
\begin{proof}
We show that for any set of equations that can potentially yield an active pair,
\lstinline;evaluate(); will generate this active pair. Such a set of equations
has the general form $\{ A=x_1,\; x_1 = x_2,\; x_2=x_3,\; \ldots,\;
x_{i-1}=x_i,\; x_i=B \}$. Applying the communication algorithm will result in at
least one \lwcom step, no matter which of the variables of each two-variable
equation is used for sorting. Hence, the size of the set of equations decreases
by at least 1 in each iteration of the loop in \lstinline;evaluate();.
After at most $i$ loops, the communication algorithm will yield $\{A=B\}$, which
can then be reduced in the subsequent interaction phase of
\lstinline;evaluate();.
\end{proof}

\paragraph*{\textbf{Remark}}
Note that the algorithm may not be able to apply all \lwcom steps to sets of
equations that potentially do \emph{not} yield an active pair. For example,
$\{ A=x_1,\; x_1 = x_2,\; x_2=x_3,\; \ldots,\; x_{i-1}=x_i,\; x_i=y \}$
may not reduce to $\{A=y\}$ depending on the choice of comparison
variables for sorting. However, these \lwcom steps do not yield any active
pairs and hence are of secondary importance, much like \lwcol and \lwsub (cf. Theorem
\ref{thm_intcom} in Section \ref{sec:lightweight_calculus}).



\section{Benchmarks and future improvements}
\label{sec:benchmarks}

In this section, we present some benchmark results and discuss possible
performance improvements. First, we compare \emph{ingpu} to existing, sequential
interaction nets evaluators. Second, we identify performance bottlenecks and
discuss potential solutions.

\subsection{Performance comparison}
\label{sec:performance-comparison}
Our implementation is still considered experimental. Unfortunately, \emph{ingpu}
currently performs slower than the more mature sequential evaluators
\emph{inets}\cite{inets_project_site} and \emph{amineLight}\cite{DBLP:journals/eceasst/HassanMS10} in most cases.
However, we can identify which parts of \emph{ingpu} are slow and which ones are fast. Consider the benchmark comparison
in Figure \ref{fig:benchmarks}. We ran our tests on a machine with an Intel Core
Duo 2.2Ghz CPU, 2GB of RAM and a Geforce GTX 460 graphics card.

\begin{figure}[hptb]
\begin{center}
\begin{center}
  \begin{tabular}{ l  c  c  r  }
\toprule
    time in seconds & amineLight & inets & ingpu \\
\midrule
    Ackermann(3,7)   & 0.4 & 0.59 & 30.4 \\
    Fibonacci(20)    & 0.03 & 0.043 & 10.5 \\
    L-System(27)     & 1.13 & 1.49 & 1.28 \\
    \bottomrule
  \end{tabular}
\end{center}
  \caption{Benchmark results}
  \label{fig:benchmarks}
\end{center}
\end{figure}

On standard benchmark functions like \emph{Ackermann} or \emph{Fibonacci},
\emph{ingpu} performs very poorly. The sequential implementations are much
faster here. However, the result on the third benchmark \emph{L-System} is much
more competitive, at least outperforming \emph{inets}. This set of rules
computes the $n$th iteration of a basic L-System that models the growth of Algae
\cite{wiki:l-system}, given by the following rewrite System:
\begin{align*}
A & \rightarrow AB & B & \rightarrow A
\end{align*}
So why is \emph{ingpu} so much faster for the \emph{L-System} ruleset?
Two main reasons can be given, which at the same time highlight the most glaring
performance problems:

\paragraph{\textbf{Inefficient communication}} Profiling
shows that for \emph{Ackermann}, almost all of the execution time of
\emph{ingpu} is spent in the communication phase and the merging part of the
interaction phase. The actual interaction kernel
(replacing equations) 
runs very fast, taking less than 0.1 percent of
the total time.

\begin{figure}[hbtp]
\begin{center}
  \includegraphics[scale=0.5]{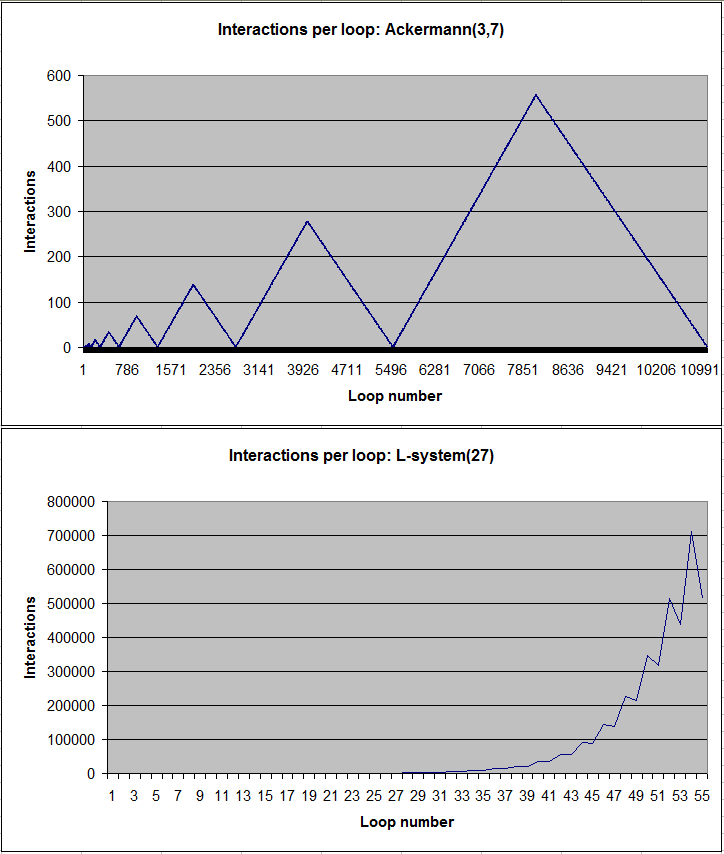}
  \caption[labelInTOC]{Available Parallelism of \emph{Ackermann} and
  \emph{L-System}: the $x$-axis shows the loop number, the $y$-axis the number
  of parallel interactions.}
  \label{fig:available_parallelism}
\end{center}
\end{figure}

\paragraph{\textbf{Available parallelism}}
The term \emph{available parallelism} was coined in
\cite{DBLP:conf/ppopp/Mendez-LojoNPSHKBP10}, and refers to the number of
expressions that can be evaluated in parallel at a given time during the
execution of a program. In our case, this is simply the number
of active pairs that exist at the same time at a certain iteration of
\emph{ingpu}'s main loop. A high number of active pairs fits the GPU
programming model well, as GPUs are optimized to handle hundreds of thousands of
data-parallel inputs. Conversely, a low amount of parallelism is not sufficient
to leverage the full computing power of the GPU. Figure
\ref{fig:available_parallelism} shows the number of parallel
interactions per loop for \emph{Ackermann} and \emph{L-System}. While
\emph{Ackerman} has less than 200 concurrent active pairs for the majority of
its execution time, \emph{L-System} performs several hundreds of thousands of
parallel interactions towards the end of the computation. Moreover,
\emph{L-System} performs much fewer loop iterations overall (50 vs. 11000). This
means that much less time is spent in the slow communication phase in
proportion to the total number of interactions.

%
Figure \ref{fig:available_parallelism} also gives
some insight on the dynamics of available parallelism in our benchmarks.
Interestingly, the number of concurrent active pairs for \emph{Ackermann} repeatedly decreases and
increases in a quasi oscillating fashion. This results in the rather low average
number of active pairs per loop. In contrast, the available parallelism in the
execution of \emph{L-System} strongly increases, as is expected considering the
exponential growth of the L-System.
The slight drops in parallelism are the result of duplicating the parameter for
the number of iterations 
in every loop.

\subsection{Possible optimizations}
\label{sec:problems-and-solutions}
Currently, \emph{ingpu} is still in its experimental stages, and
various small and big improvements can be made to increase its overall
performance. In particular, we have identified a few optimizations that have the potential for a
considerable performance increase. These are subject of current and future work.

\paragraph{\textbf{Improved communication}} As we discussed, our current
communication algorithm is very slow. Due to the fact that an interaction net is
represented as a list of independent equations, we have to sort the complete
list repeatedly to find communication-eligible equations. However, it is
possible to reduce this search space: only equations that are directly connected
to a given active pair (i.e., its interface) are potentially
communication-eligible. We gain a speedup by considering this subset of equations only. In order to
efficiently determine the eligible equations, a pointer-based net
representation in the GPU memory is needed (see the next paragraph).

\paragraph{\textbf{A more efficient net representation}}
Both interaction and communication could benefit from a better net
representation in the GPU memory. While the ``array of equations'' approach
closely follows the theoretical definition of the interaction nets calculus, it
is quite slow: the lack of a more sophisticated pointer structure makes a
sorting pass in the communication phase necessary. We are currently
implementing an approach where all agents and variables are stored in an array
such that their unique id determines their array position. This way, we achieve
constant-time access of agents and connections, making the sorting phase
unnecessary.
%

\paragraph{\textbf{Remove result array merging}} The merging of the result
vectors in the interaction phase (including dummy removal) could be improved by
using CUDA's \emph{atomic exchange} operations: these functions allow individual
kernel instances to read and write shared values. A shared pointer to the result
vector could be provided to each thread. The threads could then add the
dynamically-sized list of RHS equations to the result vector and update the
pointer to the end of their output, removing the
dummy removal pass.

A different approach to the problem of dynamic output size in GPU computations
can be found in \cite{DBLP:journals/cg/LippWW10}: the authors propose a way to
handle different output sizes without any communication between threads. The
memory management in the output array is achieved by using \emph{parallel scan
algorithms}. We are currently adapting this approach to \emph{ingpu}. Initial
experiments show that this will indeed yield a considerable speedup.


\section{Discussion}
\label{sec:discussion}

\paragraph{\textbf{Related work}}
Evaluation of interaction nets can be considered an \emph{irregular algorithm},
in the sense that it operates on a pointer structure (a graph) rather than a
dense array. Bridging the conceptual gap between the irregular nature of
interaction nets and the dense structure of typical GPU programs (e.g., dense
matrix operations) is strongly related to the implementation challenges
discussed in this paper. We have been inspired by the insights
of parallelizing irregular algorithms in \cite{DBLP:conf/ppopp/Mendez-LojoNPSHKBP10} when
implementing \emph{ingpu}. We also borrowed the term \emph{available
parallelism} from this work.

The efficient parallelization of general graph algorithms using GPUs has been
the topic of several publications (for example,
\cite{DBLP:conf/ppopp/HongKOO11}). As part of future work, we plan to use these
insights to achieve a better representation of interaction nets on the GPU.

Besides the previously mentioned \emph{inets} and \emph{amineLight},
several other interaction nets evaluators exist, for example
\cite{jose_bacelar_almeida_tool_2008,DBLP:conf/rta/Lippi02}. Another recent tool is PORGY
\cite{DBLP:journals/corr/abs-1102-2654}, which can be used to analyse and
evaluate interaction net systems with a focus on evaluation strategies.
However, only few works (for example, \cite{DBLP:conf/rta/Pinto01}) on parallel
evaluation of interaction nets exist. This is surprising, considering their
potential for parallelism. To the best of our knowledge, there is no previous
work on a GPU-based implementation.

With regard to functional programming in general, several systems based on GPUs
have been developed. Two recent examples are \emph{Obsidian}
\cite{DBLP:journals/procedia/SvenssonCS10} and \emph{Accelerate}
\cite{DBLP:conf/popl/ChakravartyKLMG11}, both being extensions of Haskell.

\paragraph{\textbf{Conclusion}}
In this paper, we have presented ongoing work on \emph{ingpu}, a
GPU-based evaluator for interaction nets. This is a novel approach that heavily
utilizes their potential for parallelism: all active pairs that are available at
the same time are evaluated in parallel. Previous evaluators are sequential or only
allow a fixed number of concurrent interactions (e.g., capped by the number of
cores of the CPU). A GPU with hundreds of cores is better suited to perform a
high number of small computations (i.e., reductions of active pairs).
Still, the implementation poses a challenge due to the dynamic nature of
interaction nets evaluation and the restrictions of the GPU computing model.

The work-in-progress status of \emph{ingpu} is clearly visible in the benchmark
results of Section \ref{sec:benchmarks}. While parallel evaluation is
generally expected to be faster than a sequential one, our current
implementation mostly performs weaker than existing evaluators. However, we
argue that the potential of \emph{ingpu} can be seen in the difference between
the individual results. In our \emph{L-System} test, \emph{ingpu} performs more
than 50 times faster than for \emph{Ackermann}, in terms of interactions per
second. The major part of the slowdown is caused by the communication phase,
which should be seen as an intermediate solution. In contrast to this, the
interaction phase (parallel reduction of the active pairs) is very fast and
shows that interaction nets and GPU are a promising match.

For current and future work, we plan to
optimize the system to improve the obvious performance bottlenecks.
Initial tests show that with a more efficient net representation in GPU memory
and the removal of result array merging (see Section
\ref{sec:problems-and-solutions}), \emph{ingpu} strongly outperforms CPU-based
systems at least for highly parallel benchmarks.

\bibliographystyle{eptcs}
\bibliography{jiresch}

\end{document}